\documentclass[aps,pra,twocolumn,reprint]{revtex4-2}
\usepackage{graphicx}
\usepackage[english]{babel}
\usepackage[inline]{enumitem}
\usepackage{amsmath, amssymb, amsfonts}
\usepackage[caption=false,labelformat=simple]{subfig}
\usepackage{bm}
\usepackage{xcolor}
\usepackage[colorlinks=true,linkcolor=blue]{hyperref}
\usepackage[normalem]{ulem}

\newlength{\figwidth}
\newlength{\lfig}
\newlength{\sfig}
\begin{document}

\title{SU($N$) symmetry with ultracold alkali dimers:\\
weak dependence of scattering properties on hyperfine state}

\author{Bijit Mukherjee}
\email{bijit.9791@gmail.com}
\thanks{Present address: Faculty of Physics, University of Warsaw, Pasteura 5, 02-093 Warsaw, Poland}
\affiliation{Joint Quantum Centre (JQC)
Durham-Newcastle, Department of Chemistry, Durham University, South Road,
Durham, DH1 3LE, United Kingdom.}
\author{Jeremy M. Hutson}
\email{j.m.hutson@durham.ac.uk} \affiliation{Joint Quantum Centre (JQC)
Durham-Newcastle, Department of Chemistry, Durham University, South Road,
Durham, DH1 3LE, United Kingdom.}

\date{\today}

\begin{abstract}
We investigate the prospect of using ultracold alkali diatomic molecules to implement many-body quantum systems with SU($N$) symmetry. Experimentally accessible molecules offer large $N$ for both bosonic and fermionic systems, with both attractive and repulsive interactions. We carry out coupled-channel scattering calculations on pairs of NaK, NaRb and NaCs molecules that are shielded from destructive collisions with static electric fields. We develop new methods to handle the very large basis sets required to include nuclear spins. We show that all the molecules studied have the properties required for SU($N$) symmetry: the collisions are principally elastic, and the scattering lengths depend only weakly on the spin states of the molecules. The rates of spin-changing inelastic collisions are very low. We develop and test a semiclassical model of the spin dependence and find that it performs well.
\end{abstract}

\maketitle

\section{Introduction}

Ultracold polar molecules offer exciting possibilities for exploring physical phenomena that range from quantum simulation \cite{Blackmore:2019,Cornish:2024} and quantum computing~\cite{DeMille:2002,Yelin:2006} to quantum magnetism~\cite{Gorshkov:2011, Gadway:2016}. To produce a stable ultracold gas in an optical trap, it is necessary to \emph{shield} pairs of molecules from close collisions that otherwise cause trap loss \cite{Ospelkaus:react:2010, Ye:2018, Gregory:RbCs-collisions:2019}. There have been various theoretical proposals to achieve shielding, with static electric fields \cite{Avdeenkov:2006, Wang:dipolar:2015, Quemener:2016, Gonzalez-Martinez:adim:2017, Mukherjee:CaF:2023, Mukherjee:alkali:2024}, near-resonant microwaves \cite{Karman:shielding:2018, Lassabliere:2018, Karman:shielding-imp:2019} or optical fields \cite{Xie:optical:2020}. Experiments have demonstrated the efficiency of both static-field and microwave shielding for various molecules \cite{Matsuda:2020, Li:KRb-shield-3D:2021, Anderegg:2021, Bigagli:NaCs:2023, Lin:NaRb:2023}. Notably, microwave shielding has led to the achievement of Fermi degeneracy for Na$^{40}$K \cite{Schindewolf:NaK-degen:2022} and Bose-Einstein condensation for NaCs \cite{Bigagli:BEC:NaCs:2024}.

We recently showed \cite{Mukherjee:SU(N):2025} that shielded ultracold molecules can exhibit many-body properties associated with SU($N$) symmetry, where $N$ is the number of available spin states. This will open up exciting possibilities for studying novel aspects of quantum magnetism. To realize SU($N$) symmetry, the interactions must be dominated by elastic scattering and independent of the states involved. In ultracold systems, SU($N$) symmetry has been predicted and realized with nuclear spin states of alkaline-earth-like atoms (Sr and Yb), which allow $N$ up to 10 \cite{Gorshkov:SU(N):2010,Cazalilla:2014}. However, the experimentally viable alkaline-earth-like atoms with nuclear spin are all fermions and have repulsive interactions, i.e., their scattering lengths $a$ are positive. Shielded molecules, on the other hand, can be either bosonic or fermionic, with scattering lengths that can be either positive or negative \cite{Mukherjee:alkali:2024,Lassabliere:2018}. They also possess substantial electric dipole moments. The experimentally accessible bialkali molecules (hereafter ``alkali dimers'') might exhibit SU($N$) symmetry with $N$ as large as 36. The high symmetry would enhance quantum fluctuations and stabilize exotic states of matter such as chiral spin liquids \cite{Hermele:2009,Chen:multiflavor:2024}. It would also allow the study of dynamical phenonema such as prethermalization~\cite{Huang:suppression:2020} in nonequilibrium quantum systems.

Deviations from SU($N$) symmetry can be characterized by the range of scattering lengths for colliding pairs in different states. In ref.\ \cite{Mukherjee:SU(N):2025}, we carried out coupled-channel scattering calculations for CaF molecules in different combinations of spin states, shielded with static electric fields. We showed that the rates of spin-changing collisions are very low and that the deviations $\delta\alpha$ in the real part of the scattering length from its spin-free value $\alpha_0$ are only about 3\%. To explain the values of $\delta\alpha$, we developed a model based on the long-range adiabatic potential corresponding to the shielded interaction of a pair of molecules. The model gave good agreement with the coupled-channel results including electron and nuclear spins. It also predicted relative deviations $\delta\alpha/\alpha_0$ for most of the shielded alkali dimers that are even smaller than for CaF.

In this paper, we study the dependence of the scattering properties on hyperfine state for pairs of alkali dimers shielded with electric fields. We carry out coupled-channel scattering calculations on shielded collisions, taking account of hyperfine structure. We test the model of spin-dependence developed in ref.\ \cite{Mukherjee:SU(N):2025} and show that it gives reliable predictions. We also investigate collisions that can change the hyperfine state of shielded molecules. These spin-changing collisions are slow, but not negligible.

A major challenge in this work is the size of the basis sets involved. The basis-set size scales approximately with $N^2$. The computer time scales as the cube of this, so as $N^6$. For the alkali dimers, $N=(2i_\textrm{A}+1)(2i_\textrm{B}+1)$, where $i_\textrm{A}$ and $i_\textrm{A}$ are the nuclear spins of the two atoms. This can be as large as 36, compared to 4 for CaF. Even for CaF, calculations with a full spin basis take hours of computer time for a single value of electric field. Without special techniques, calculations for $N=36$ would take decades or more. To circumvent this, we develop techniques with restricted spin bases and demonstrate their accuracy.

The structure of the paper is as follows. Section \ref{sec:methods} describes our coupled-channel approach, including the Hamiltonians and basis sets. Section \ref{sec:results} presents our results on the spin dependence of scattering lengths and the rates of spin-changing collisions for alkali dimers of current interest. It describes the mechanisms of the spin-changing processes and demonstrates that a restricted spin basis set reproduces the scattering properties accurately. It demonstrates that the semiclassical model of spin dependence performs well for these systems. Finally, Section \ref{sec:conclusions} presents conclusions and perspectives.

\section{Methods}
\label{sec:methods}

\subsection{Coupled-channel approach}
\label{ssec:cc}

The theory has been described in detail in ref.\ \cite{Mukherjee:CaF:2023}, and only a brief summary will be given here to define notation.

In the presence of an external static electric field $\boldsymbol{F}$, the Hamiltonian for a spin-free diatomic molecule $k$ in a $^1\Sigma^+$ state in the rigid-rotor approximation is
\begin{equation}
\hat{h}_k = b_k\hat{\boldsymbol{n}}_k^2 - \boldsymbol{\mu}_k \cdot \boldsymbol{F},
\label{eq:ham-mon-spin-free}
\end{equation}
where $\hat{\boldsymbol{n}}$ is the operator for molecular rotation, $b$ is the rotational constant and $\boldsymbol{\mu}$ is the dipole moment along the molecular axis. The values of the molecular constants are tabulated in ref.\ \cite{Mukherjee:alkali:2024}.

For a pair of colliding molecules the Hamiltonian is
\begin{equation}
\hat{H} = \frac{\hbar^2}{2\mu_\textrm{red}}\left( -R^{-1} \frac{d^2}{dR^2} R  + \frac{\hat{\boldsymbol{L}}^2}{R^2} \right)
+ \hat{H}_\textrm{intl} + V_\textrm{int},
\label{eq:ham-pair}
\end{equation}
where $R$ is the intermolecular distance, $\hat{\boldsymbol{L}}$ is the angular momentum operator for relative rotation of the two molecules and $\mu_\textrm{red}$ is the reduced mass. The internal Hamiltonian $\hat{H}_\textrm{intl}$ is
\begin{equation}
\hat{H}_\textrm{intl} = \hat{h}_1 + \hat{h}_2
\label{eq:ham-int}
\end{equation}
and $V_\textrm{int}$ is the interaction potential. Since the interactions involved in shielding occur at long range, we approximate $V_\textrm{int}$ by the dipole-dipole operator,
\begin{equation}
\hat{H}_\textrm{dd} = -[3(\boldsymbol{\mu}_1\cdot\hat{\boldsymbol{R}}) (\boldsymbol{\mu}_2\cdot\hat{\boldsymbol{R}}) - \boldsymbol{\mu}_1\cdot\boldsymbol{\mu}_2] / (4\pi\epsilon_0 R^3),
\label{eq:Hdd}
\end{equation}
where $\hat{\boldsymbol{R}}$ is a unit vector along the intermolecular axis.
This is supplemented by an electronic dispersion term $V^\textrm{elec}_\textrm{disp}=-C^\textrm{elec}_6/R^6$, with values of $C^\textrm{elec}_6$ from ref.\ \cite{Lepers:2013}.

The total wavefunction is expanded
\begin{equation} \Psi(R,\hat{\boldsymbol{R}},\hat{\boldsymbol{r}}_1,\hat{\boldsymbol{r}}_2)
=R^{-1}\sum_j\Phi_j(\hat{\boldsymbol{R}},\hat{\boldsymbol{r}}_1,\hat{\boldsymbol{r}}_2)\psi_{j}(R), \label{eq:expand}
\end{equation}
where $\hat{\boldsymbol{r}}_k$ is a unit vector along the axis of molecule $k$.
We use a basis set of functions $\{\Phi_j\}$,
\begin{equation}
\Phi_j = \phi^{\tilde{n}_1}_{m_{n1}}(\hat{\boldsymbol{r}}_1) \phi^{\tilde{n}_2}_{m_{n2}}(\hat{\boldsymbol{r}}_2) Y_{LM_L}(\hat{\boldsymbol{R}}),
\label{eq:basis-spin-free}
\end{equation}
symmetrized for exchange of identical molecules. Here $\phi^{\tilde{n}_1}_{m_{n1}}(\hat{\boldsymbol{r}}_1)$ and $\phi^{\tilde{n}_2}_{m_{n2}}(\hat{\boldsymbol{r}}_2)$ are field-dressed rotor functions that diagonalize $\hat{h}_1$ and $\hat{h}_2$, respectively, and $Y_{LM_L}(\hat{\boldsymbol{R}})$ are spherical harmonics that are the eigenfunctions of $\hat{\boldsymbol{L}}^2$. Here $\tilde{n}$ is a hindered-rotor quantum number that correlates with the free-rotor quantum number $n$ at zero field and $m_n$ represents the conserved projection of $n$ onto the space-fixed $z$ axis, chosen to lie along the static electric field. The field-dressed functions $\phi^{\tilde{n}}_{m_{n}}(\hat{\boldsymbol{r}})$ are expanded in free-rotor functions $Y_{nm_n}(\hat{\boldsymbol{r}})$.

Substituting the expansion (\ref{eq:expand}) into the total Schr\"odinger equation produces a set of coupled equations in the intermolecular distance $R$. The eigenvalues of $\hat{H}_\textrm{intl} + V_\textrm{int} + \hbar^2\hat{\boldsymbol{L}}^2/2\mu_\textrm{red}R^2$ form a set of adiabats that represent effective potentials for relative motion of the two molecules. To a first approximation, collisions may be viewed as taking place on these adiabats, although transitions between them are fully included in coupled-channel calculations.

Colliding molecules may be lost from a trap in two ways. First, a colliding pair may undergo a transition to a lower-lying pair state. Such inelastic collisions release kinetic energy that is usually sufficient
to eject both molecules from the trap. Secondly, a colliding pair that reaches small intermolecular distance is likely to be lost through processes that may include short-range inelasticity, laser absorption, or a three-body collision. To model these processes, we solve the coupled equations subject to a fully absorbing boundary condition at short range \cite{Clary:1987, Janssen:PhD:2012}. The numerical methods used are as described in ref.\ \cite{Mukherjee:CaF:2023}. This is done separately for each electric field $F$ and collision energy $E_\textrm{coll}$, producing a nonunitary S matrix and cross sections for elastic scattering, state-to-state inelasticity and total collisional loss. The corresponding rate coefficients $k$ are related to the cross sections $\sigma$ through $k = v\sigma$, where $v =
(2E_\textrm{coll}/\mu_\textrm{red})^{1/2}$. We also extract the complex energy-dependent scattering length \cite{Hutson:res:2007}, whose real part characterizes the overall strength of the interaction.

Shielding is achieved by engineering a long-range repulsive interaction that prevents colliding pairs reaching the short-range region where most loss occurs. This can happen at electric fields where the incoming pair state lies just above another pair state that is connected to it by $\hat{H}_\textrm{dd}$. The repulsion due to mixing of the two pair states creates the shielding barrier. All calculations in the present paper are for molecules in their ground vibronic state (X$^1\Sigma^+$, $v=0$), with initial rotor quantum numbers $(\tilde{n},m_n)=(1,0)+(1,0)$, at electric fields just above the crossing with $(0,0)+(2,0)$. We use a collision energy $E_\textrm{coll} = 10$ nK$\times k_\textrm{B}$, which is a reasonable lower limit for the temperatures that are likely to be experimentally accessible and where quantum degeneracy may be attained.

\subsection{Inclusion of hyperfine structure}

Alkali-metal atoms possess nuclear spins $i$ which interact with one another and with rotation to produce hyperfine splittings. For $i>1/2$ and $n\ne0$, the largest interaction is usually the nuclear quadrupole coupling, which arises due to the nuclear electric quadrupole moment $ e\boldsymbol{Q}$ interacting with the electric field gradient $\boldsymbol{q}$ in the molecule. The hyperfine Hamiltonian $\hat{h}_\textrm{hf}$ for a molecule AB in $^1\Sigma^+$ state is
\begin{eqnarray}
\hat{h}_\textrm{hf} &=& \sum_{\alpha=\textrm{A,B}} e\boldsymbol{Q}_{\alpha} \cdot \boldsymbol{q}_{\alpha}
 + \sum_{\alpha=\textrm{A,B}} c_{\alpha} \boldsymbol{n} \cdot \boldsymbol{i}_{\alpha} \nonumber \\
&\,& - c_3 \sqrt{6} T^2(C) \cdot T^2(\boldsymbol{i}_\textrm{A}, \boldsymbol{i}_\textrm{B}) +
c_4 \boldsymbol{i}_\textrm{A} \cdot \boldsymbol{i}_\textrm{B}.
\label{eq:ham-mon-hf}
\end{eqnarray}
Here $\boldsymbol{i}_\textrm{A}$ and $\boldsymbol{i}_\textrm{B}$ are the spins of nuclei A and B. The first term is the nuclear quadrupole interaction, characterized by the coupling constants $(eQq)_\textrm{A}$ and $(eQq)_\textrm{B}$. The second is the interaction between the nuclear spins and the rotation of the molecule, characterized by the spin-rotation coupling constants $c_\textrm{A}$ and $c_\textrm{B}$. The third term represents the anisotropic interaction between the two nuclear spins, characterized by the coupling constant $c_3$; $T^2(\boldsymbol{i}_\textrm{A}, \boldsymbol{i}_\textrm{B})$ denotes a rank-2 spherical tensor formed from $\boldsymbol{i}_\textrm{A}$ and $\boldsymbol{i}_\textrm{B}$, and $T^2(C)$ is a spherical tensor whose components are the Racah-normalized spherical harmonics $C^2_q(\theta,\phi)$. The last term is the isotropic spin-spin interaction, with coupling constant $c_4$. We use hyperfine constants from refs.\ \cite{Aldegunde:polar:2008, Aldegunde:singlet:2017, Gregory:RbCs-microwave:2016, Guo:NaRb:2018}. The spin-rotation and spin-spin couplings are typically more than 100 times smaller than the quadrupole coupling. We add the hyperfine Hamiltonian $\hat{h}_\textrm{hf}$ to the single-molecule Hamiltonian of Eq.\ \ref{eq:ham-mon-spin-free} to form the internal Hamiltonian $\hat{H}_\textrm{intl}$ in Eq.\ \ref{eq:ham-int}.

\subsection{Basis sets}
\label{sec:basis}

We construct basis sets from products of field-dressed rotor functions $|\tilde{n},m_n\rangle$ and spin functions $|(i_\textrm{A}, i_\textrm{B})i, m_i\rangle$. Here $i$ is the resultant of the two nuclear spins $i_\textrm{A}$ and $i_\textrm{B}$ and $m_i=m_{i,\textrm{A}}+m_{i,\textrm{B}}$, where $m_{i,\textrm{A}}$ and $m_{i,\textrm{B}}$ are the projections of $i_\textrm{A}$ and $i_\textrm{B}$ along the $z$ axis. The products are symmetrized for exchange. We multiply these by functions for the partial-wave quantum number $L$ and its projection $M_L$.

We include field-dressed rotor functions up to $\tilde{n}_\textrm{max}=5$. However, this gives a basis set too large to be used directly in coupled-channel calculations, particularly when spins are included. We therefore divide the basis functions into two groups, namely ``class 1'' and ``class 2'', according to the pairs of rotor functions involved. The class 1 pair functions are used explicitly in the coupled-channel calculations, while the class 2 functions are taken into account through Van Vleck transformations as described in ref.\ \cite{Mukherjee:CaF:2023}. For spin-free calculations, we include all functions with $\tilde{n}\le2$ in class 1. To allow the inclusion of spin functions, we restrict this further and include only $N_\textrm{rot}=14$ rotor pairs in class 1, chosen to be those that are closest in energy to the incoming channel \cite{Mukherjee:SU(N):2025}.

Except for molecules containing $^6$Li, the alkali dimers with the smallest $N$ are those where both nuclei have $i=3/2$, such as $^{23}$Na$^{39}$K and $^{23}$Na$^{87}$Rb. We consider $^{23}$Na$^{39}$K as a prototype system to illustrate the sizes of the basis sets for coupled-channel calculations including spin, both to present calculations including full hyperfine structure and to discuss the approximations that can be made. For $^{23}$Na$^{39}$K, inclusion of all spin functions for every rotor pair up to $\tilde{n}_\textrm{max}=5$ gives $N_\textrm{pair}\approx 10^5$ pair functions for each $L$, $M_L$. The Van Vleck transformation with $N_\textrm{rot}=14$ reduces this to $N_\textrm{pair}=3224$ in class 1.

For spin-free calculations, we include all functions for partial waves $L$ up to $L_\textrm{max}=20$. When spin is included, we restrict $L_\textrm{max}$ to 4 or 6. This gives good convergence for the spin dependence, though not for spin-free properties \cite{Mukherjee:SU(N):2025}. As in refs.\ \cite{Mukherjee:CaF:2023, Mukherjee:SU(N):2025}, we name the resulting basis sets spin-N$\langle N_\textrm{pair} \rangle$-L$ \langle L_\textrm{max} \rangle$. In the presence of an external field, the projection of the total angular momentum for the colliding pair, $M_\textrm{tot}=m_{n,1}+m_{i,1}+m_{n,2}+m_{i,2}+M_L$ is conserved. We perform calculations only for the values of $M_\textrm{tot}$ for which the s-wave incoming channel is included in the basis set (or p-wave for identical fermions). This is a reasonable approximation at the low collision energies considered here.

\begin{figure}[tbp]
\begin{center}
	\includegraphics[width=0.45\textwidth]{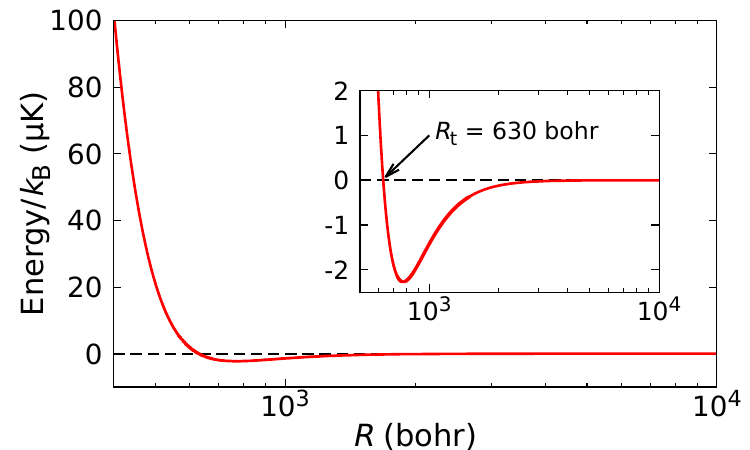}
    \caption{The adiabat that correlates asymptotically with the s-wave channel of the incoming rotor pair $(\tilde{n},m_n)=(1,0)+(1,0)$ for Na$^{39}$K at $F=7.1$ kV/cm. The repulsive wall is responsible for shielding. The inset shows an expanded view of the long-range well, with the position $R_\textrm{t}$ of the inner turning point indicated.}%
    \label{fig:adiabat}
\end{center}
\end{figure}

\subsection{Semiclassical model}
\label{sec:model}
Ref.\ \cite{Mukherjee:SU(N):2025} developed a semiclassical model of $\delta\alpha(F)$, based on the effective potential curves (adiabats) that govern shielding. The model is expressed in terms of a phase integral $\Phi$ over the lowest adiabat that correlates with the initial pair state, together with the zero-energy inner turning point $R_\textrm{t}$ for this adiabat. An example of such an adiabat is shown in Fig.\ \ref{fig:adiabat}, for the case of Na$^{39}$K at $F=7.1$ kV/cm. The matrix elements of $\hat{H}_\textrm{dd}$ off-diagonal in $L$ produce a long-range attraction that is asymptotically proportional to $1/R^4$ \cite{Bohn:BCT:2009}. In a single-channel model, the scattering length $a$ for such a potential is
\begin{equation}
a = R_\textrm{t} - \sqrt{8/15} D \tan\left(\Phi-\frac{\pi}{4}\right),
\label{eq:a-Phi}
\end{equation}
where $D=d_1 d_2 \mu_\textrm{red}/(4\pi\epsilon_0\hbar^2)$ is the dipole length for interaction of two molecules with space-fixed dipoles $d_1$ and $d_2$. The first term was omitted in ref.\ \cite{Gribakin:1993}, but arises simply because all semiclassical wavefunctions are expressed with respect to an origin at the classical turning point. This term is important for shielded molecules, where $R_\textrm{t}$ can be very large; it accounts for the excluded volume due to the repulsive part of the shielding potential. The term involving $\Phi$ accounts for the attractive potential well at long range.

Both $R_\textrm{t}(F)$ and $\Phi(F)$ are functions of field $F$. Their spin-free values $R_{\textrm{t}0}(F)$ and $\Phi_0(F)$, together with the corresponding scattering length $a_0(F)$, may be obtained from spin-free calculations on a colliding pair of molecules, which are computationally far cheaper than calculations including spin structure. $\Phi_0$ may be obtained either by integration over the lowest adiabat or from the real part of the scattering length, using Eq.\ \ref{eq:a-Phi}. The two approaches give similar results when shielding is effective; in the present work, we use the latter.

We showed in ref.\ \cite{Mukherjee:SU(N):2025} that, for CaF at fixed field, $R_\textrm{t}$ and $\Phi$ for different spin combinations have simple dependences on the space-fixed dipoles of the colliding molecules. Specifically,
\begin{equation}
R_\textrm{t}(F) \propto [D(F)]^{-1} \hbox{\quad and \quad} \Phi(F)\propto[D(F)]^2.
\end{equation}
This allows differentiation of Eq.\ \ref{eq:a-Phi} to obtain
\begin{align}
\frac{da}{dD}
\approx (a-2R_\textrm{t})/D - 2 \Phi \sqrt{8/15} \sec^2\left(\Phi-\frac{\pi}{4}\right).
\label{eq:dabydD}
\end{align}
The scattering length for collision between a pair of molecules in spin states $j$ and $j'$, with space-fixed dipoles $d_j$ and $d_{j'}$, may be expanded about the corresponding spin-free value at field $F$. The deviations $\delta\alpha_{jj'}$ of the real parts $\alpha_{jj'}$ of the scattering lengths from the spin-free value $\alpha_0$ are
\begin{equation}
\delta\alpha_{jj'}= (D_{jj'}-D_0) \frac{da}{dD} \approx D_0 \left(\Delta d_j + \Delta d_{j'}\right) \frac{da}{dD},
\label{eq:Delta-alpha}
\end{equation}
where $D_{jj'}=d_j d_{j'} \mu_\textrm{red}/(4\pi\epsilon_0\hbar^2)$ and $D_0$ is the corresponding spin-free value. $\Delta d_j=(d_j-d_0)/d_0$ is the fractional change in the space-fixed dipole moment for spin state $j$ from the spin-free value $d_0$. All these quantities are readily evaluated from calculations on individual molecules, without scattering calculations that explicitly include spin.

In this paper we test this model for the alkali dimers by comparing its results with those of coupled-channel calculations including nuclear spin. The model may be viewed as involving two approximations: first that the deviations $\delta\alpha_{jj'}$ are proportional to $\left(\Delta d_j + \Delta d_{j'}\right)$, as in Eq.\ \ref{eq:Delta-alpha}, and second that the constant of proportionality is given by Eq.\ \ref{eq:dabydD}. We find that the first approximation is very accurate, while the second can lose accuracy when there is strong cancellation between the terms arising from the excluded volume and the potential well.

\section{Results}
\label{sec:results}

\subsection{Hyperfine levels of shielded alkali dimers}

For an isolated molecule, the projection $m_f=m_n+m_i$ of the total angular momentum is conserved. In the presence of a strong electric field, $m_n$ is approximately a good quantum number and so is $m_i$. However, $i$ and the individual nuclear spins $m_{i,\textrm{A}}$ and $m_{i,\textrm{B}}$ are not good quantum numbers due to the spin-spin and nuclear quadrupole interactions.

\begin{figure}[tbp]
\begin{center}
	\includegraphics[width=0.5\textwidth]{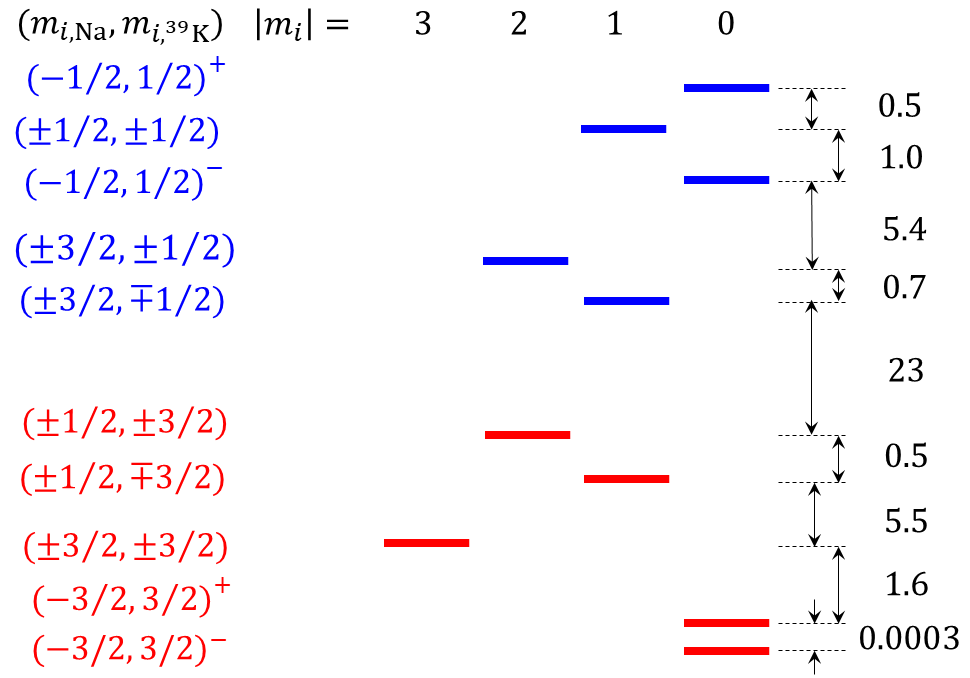}
    \caption{Hyperfine energy levels (in kHz) of Na$^{39}$K for the rotor level $(\tilde{n},m_n)$ = (1,0) at $F=7.1$ kV/cm. The states are labeled with quantum numbers $(m_{i,\textrm{Na}}, m_{i,^{39}\textrm{K}})$ and arranged in columns according to $|m_i|$, where $m_i=m_{i,\textrm{Na}}+m_{i,^{39}\textrm{K}}$. The spacings are not shown to scale.}%
    \label{fig:Alk-energy-levels}
\end{center}
\end{figure}

Figure \ref{fig:Alk-energy-levels} shows the hyperfine level structure of $^{23}$Na$^{39}$K (hereafter Na$^{39}$K) for $(\tilde{n},m_n)$ = (1,0) in the presence of an electric field $F=7.1$ kV/cm, which gives very effective shielding. The hyperfine splittings are caused mostly by the quadrupole couplings of the two nuclei. The levels are approximately split into two groups, shown in red and blue, containing 8 states each. This separation arises because $(eQq)_{^{39}\textrm{K}}$ is a factor of 4.5 larger than $(eQq)_\textrm{Na}$. The splitting between the two groups is caused by $(eQq)_{^{39}\textrm{K}}$ whereas the splittings within each group are caused by $(eQq)_\textrm{Na}$. This enables us to assign the states with approximately good quantum numbers for the nuclear spin projections $(m_{i,\textrm{Na}}, m_{i,^{39}\textrm{K}})$. The group of states shown in red (blue) have $|m_{i,^{39}\textrm{K}}|=3/2$ ($1/2$). All the states have $m_f=m_i$, since $m_n=0$. States with $m_i=0$ are symmetric and antisymmetric linear combinations of states $(m,-m)$ and $(-m,m)$; this symmetry is indicated by a superscript $+$ or $-$.

The alkali dimers have a large number of available hyperfine states, so it is impracticable to show results for all possible combinations of them. From the point of view of SU($N$) symmetry, the most important quantity is the variation of the real part $\alpha$ of the scattering length between these combinations. As in ref.\ \cite{Mukherjee:SU(N):2025}, we characterize this by $\delta\alpha_{jj'}=\alpha_{jj'}-\alpha_0$, where $\alpha_{jj'}$ is the (field-dependent) scattering length for a collision between molecules in hyperfine states $j$ and $j'$ and $\alpha_0$ is the corresponding value when hyperfine structure is neglected. Our goal is to find the \emph{range} of $\alpha_{jj'}$ between different combinations, since this characterizes deviations from SU($N$) symmetry. The model of Sec.\ \ref{sec:model} predicts that the largest (smallest) values of $\alpha_{jj'}$ will occur for collisions between identical pairs of molecules in hyperfine states with the smallest (largest) space-fixed dipole moment $d_j$ at the field concerned. This is because higher $d_j$ gives increased attraction in the long-range potential well and hence produces lower $\alpha_{jj'}$. For each molecule of interest, we diagonalize the monomer Hamiltonian at an electric field where shielding is effective, and calculate values of $d_j$. Tabulations of the resulting hyperfine levels and values of $\Delta d_j$ are given in Supplemental Material \cite{sup-mat-AlkSU_N}. For each molecule, we select the states with the largest and smallest values of $d_j$ and focus on collisions involving these two states in the results presented below.

SU($N$) symmetry also requires that collisions that convert molecules between spin states are very slow.
In alkali dimers, such transitions are caused mostly by quadrupole couplings. These are off-diagonal in ($\tilde{n},m_n$) and in $m_i$, while conserving $m_f$. In strong electric fields, however, states with different values of $|m_n|$ are far apart, even if they have the same $\tilde{n}$. Because of this, inelastic transitions due to quadrupole couplings are relatively weak. In the calculations of spin-changing rates below, we focus on collisions of pairs of molecules initially in the highest hyperfine state with $m_f=m_i=0$. Such collisions have a wide variety of energetically accessible inelastic channels, and we expect the rates of spin-changing collisions to be representative of the fastest that can occur.

\subsection{Na$^{39}$K}
\label{sec:proto}

\begin{figure}[tbp]
\begin{center}
	\includegraphics[width=0.5\textwidth]{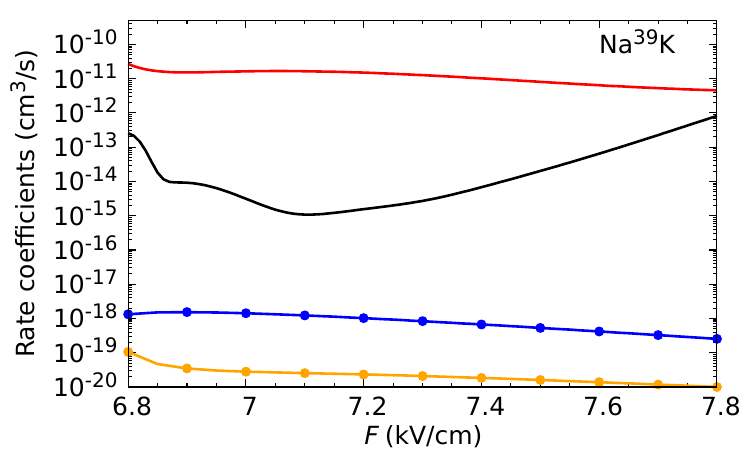}
    \caption{Rate coefficients for elastic scattering (red) and total loss (black) from spin-free coupled-channel calculations on Na$^{39}$K at collision energy $E_\textrm{coll} = 10$ nK $\times k_\textrm{B}$. The blue and orange curves show state-to-state inelastic rate coefficients for 1-molecule spin-changing inelastic transitions from initial state $(m_{i,\textrm{Na}}, m_{i,^{39}\textrm{K}})$ = $(-1/2,1/2)^+$ to $(-1/2,3/2)$ and $(-3/2,3/2)^+$, respectively, within the same rotor level. For spin-changing collisions, the lines and circles are obtained from calculations using basis sets spin-N3224-L4 and MFR1, respectively.}%
    \label{fig:Na39K:rates}
\end{center}
\end{figure}

In this section, we discuss the collision properties for two Na$^{39}$K molecules in the presence of static-field shielding. We focus on the effects of hyperfine structure, but begin by summarizing the results of spin-free calculations. The red and black lines in Fig.\ \ref{fig:Na39K:rates} show the rate coefficients for elastic scattering and total loss, calculated with the basis set of ref.\ \cite{Mukherjee:alkali:2024}, with $L_\textrm{max}=20$. Shielding is effective at fields between 6.8 and 7.4 kV/cm.

To investigate the effects of hyperfine structure, we need to use much smaller basis sets of rotor functions and much smaller $L_\textrm{max}$. As described in Sec.\ \ref{sec:basis}, we include 14 rotor pairs in class 1. When combined with all possible spin functions of Na$^{39}$K, this gives 3224 pair functions. We use $L_\textrm{max} = 4$, so that the basis set is called spin-N3224-L4; it contains a total of 4812 functions. We carry out coupled-channel calculations for different initial spin states as a function of static electric field to study $\delta\alpha_{jj'}$ and the rates of spin-changing inelastic collisions.

\begin{figure}[tbp]
\includegraphics[width=\columnwidth]{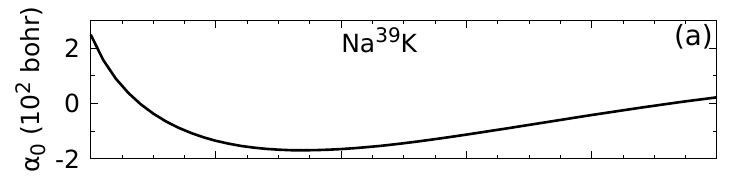}
\includegraphics[width=\columnwidth]{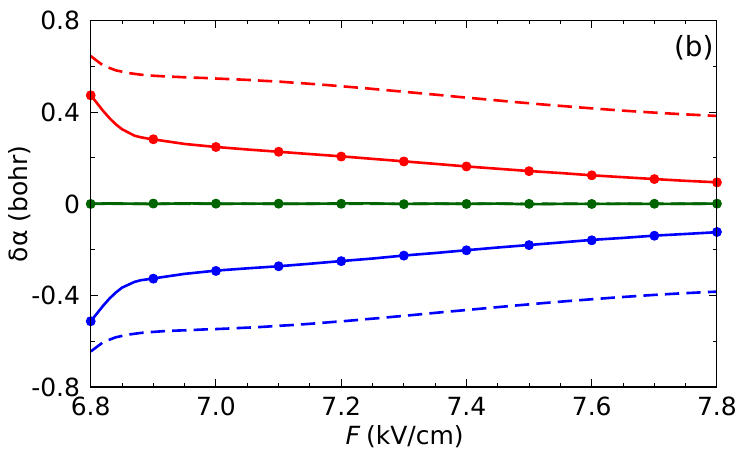}
    \caption{ (a) Real part $\alpha_0(F)$ of the scattering length for Na$^{39}$K from spin-free calculations using the basis set with $N_\textrm{rot}=14$. (b) Deviations $\delta\alpha(F)$ in scattering lengths from the spin-free value for the combinations of spin states $j+j$ (blue), $j^{\prime}+j^{\prime}$ (red) and $j+j^{\prime}$ (green), where $j=(m_{i,\textrm{Na}}, m_{i,^{39}\textrm{K}})=(-3/2,3/2)^+$ and $j^{\prime}=(-1/2,1/2)^-$. The state $j$ and $j^{\prime}$ are those with the highest and lowest space-fixed dipole moment, respectively. The solid lines and circles are obtained from coupled-channel calculations using basis sets spin-N3224-L4 and MFR1, respectively. The dashed lines are calculated using the semiclassical model of Sec.\ \ref{sec:model}.}%
    \label{fig:Na39K_diff-a}
\end{figure}

For Na$^{39}$K, the spin states with the largest and smallest values of $d_j$ are those with $(m_{i,\textrm{Na}}, m_{i,^{39}\textrm{K}})^\pm=(-3/2,3/2)^+$ and $(-1/2,1/2)^-$. The spin-free scattering length $\alpha_0(F)$ is shown in Fig.\ \ref{fig:Na39K_diff-a}(a) and the deviations $\delta\alpha_{jj'}(F)$ for collisions involving these two states are shown by the solid lines in Fig.\ \ref{fig:Na39K_diff-a}(b). The deviations are approximately equal and opposite for these two spin states, and close to zero for their combination. This is expected from Eq.\ \ref{eq:Delta-alpha}, because the two states have almost equal and opposite values of $\Delta d_j$. We have confirmed that $\delta\alpha_{jj'}$ (at fixed field) is accurately proportional to $\Delta d_j+\Delta d_{j'}$ for any combination $j+j^{\prime}$. All remaining spin combinations produce $\delta\alpha_{jj'}$ within the bounds set by the red and blue curves. The results show that scattering lengths are independent of spin state to within about 0.25\% at $F=7.1$ kV/cm, where shielding is most effective.

The values of $\delta\alpha_{jj'}$ predicted by the model are shown by the dashed lines in Fig.\ \ref{fig:Na39K_diff-a}(b). It may be seen that the model overestimates the actual deviations in $\alpha$ by about a factor of 2. The error is relatively large for NaK because, for this system, the attractive and repulsive contributions to Eq.\ \ref{eq:Delta-alpha} almost cancel, amplifying the error. As will be seen below, the model is much more accurate for alkali dimers with larger values of $D_0$.

We next study the rates of spin-changing collisions in Na$^{39}$K and their effects on the possible realization of SU($N$) symmetry. Among the various possible spin-changing inelastic transitions, those with $\Delta m_i=0$ and $\pm 1$ dominate. The orange and blue lines in Fig.\ \ref{fig:Na39K:rates} show the state-to-state rate coefficients for the fastest of each of these two types of transition for Na$^{39}$K molecules initially in their highest hyperfine level, $(-1/2,1/2)^+$, for $\Delta m_i=0$ and $\pm1$, respectively. These are essentially 1-molecule transitions with outgoing spin states $(-1/2, 3/2)$ (blue) and $(-3/2, 3/2)^+$ (orange).

The main mechanism that causes spin-changing transitions involves second-order couplings via different rotor pair states, involving both the dipole-dipole and nuclear quadrupole operators. The quadrupole operators can change $\tilde{n}$, $m_n$ and $m_i$ but conserve $m_f$, while the dipole-dipole operator conserves $m_i$ but has selection rule $\Delta m_f=\Delta m_n=0,\pm1$. The overall second-order transition can thus have $\Delta m_f=\Delta m_i=0,\pm1$. When $m_f=m_f'=0$, only transitions $+\leftrightarrow +$ or $-\leftrightarrow -$ are allowed.

For Na$^{39}$K, $(eQq)_\textrm{B}\gg (eQq)_\textrm{A}$, so that $m_{i,\textrm{A}}$ and $m_{i,\textrm{B}}$ are nearly conserved. Under the influence of single-nucleus operators such as the nuclear quadrupole coupling, only one of them can change, so $m_i$ must change. The spin-changing collisions are dominated by the incoming s wave, with $L=0$. Because of the presence of the dipole-dipole operator, the outgoing channel must have $L=2$. The transitions occur principally at long-range avoided crossings between the incoming and outgoing adiabats; for Na$^{39}$K, the dominant transitions occur at distances greater than 3000 bohr. For a fixed coupling proportional to $R^{-3}$, the rate of such transitions is proportional $\Delta E^{1/2}$, where $\Delta E$ is the kinetic energy release \cite{Kajita:04a}. This explains why the fastest transition for each $\Delta m_i$ is to the lowest state allowed by the selection rules.

For Na$^{39}$K, the total loss is dominated by inelastic transitions that change rotor quantum numbers and by short-range processes, rather than by spin-changing collisions. This is not true for all alkali dimers, as discussed below. Nevertheless, the rate of spin-changing collisions is always small enough to meet the requirements for SU($N$) symmetry.

\subsection{Restricted basis set}
\label{sec:reduced}

Even with Van Vleck transformations, the number of basis functions is extremely large when we include all hyperfine levels, particularly for molecules with individual nuclear spins $i_\alpha>3/2$. As described in subsection \ref{sec:proto}, we find that inelastic transitions are dominated by $\Delta m_f=0$ or $\pm 1$. Based on this fact, we remove all spin functions from our basis sets except those with $|m_f-m_{f,\textrm{init}}| \leq 1$, where $m_{f,\textrm{init}}$ is the value for the initial state of the colliding molecule. This substantially reduces the basis-set size: for $m_{f,\textrm{init}}=0$, $N_\textrm{pair}$ is reduced from 3224 to 1110 for Na$^{39}$K and NaRb, and from 12848 to 1818 for NaCs.

The $m_f$-restricted basis set, hereafter referred as MFR1, accurately reproduces all scattering results obtained from full spin basis sets. Figs.\ \ref{fig:Na39K:rates} and \ref{fig:Na39K_diff-a} compare results using MFR1 (circles) with those of full spin basis sets (solid lines). In the remainder of the paper, we use MFR1 with $L_\textrm{max}=6$ to carry out coupled-channel calculations including spins.

\subsection{Na$^{87}$Rb}

\begin{figure}[tbp]
\includegraphics[width=\columnwidth]{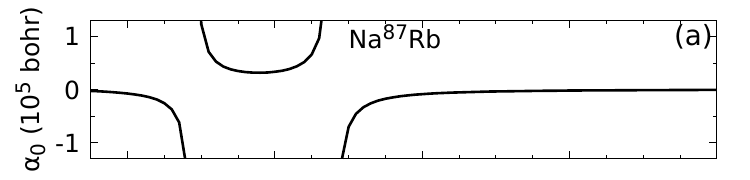}
\includegraphics[width=\columnwidth]{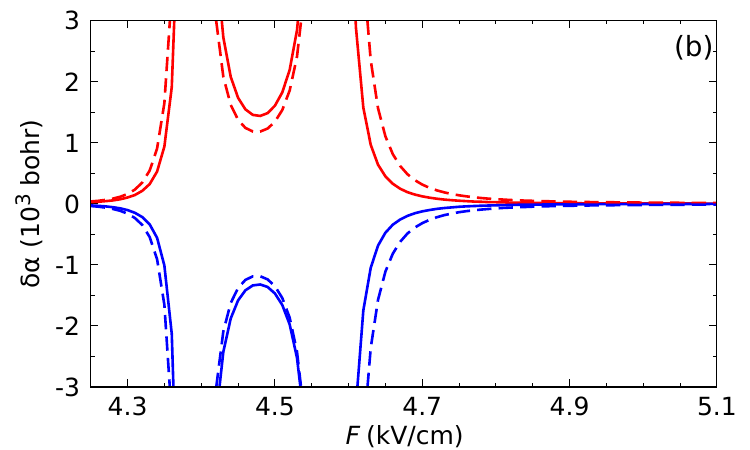}
    \caption{ (a) Real part $\alpha_0(F)$ of the scattering length for NaRb from spin-free calculations. (b) Deviations $\delta\alpha(F)$ from the spin-free value for the combinations of spin states $j+j$ (blue) and $j'+j'$ (red), where $j=(m_{i,\textrm{Na}}, m_{i,\textrm{Rb}})=(\pm3/2, \pm3/2)$ and $j'=(-1/2,1/2)^+$. The states $j$ and $j'$ are those with the highest and lowest space-fixed dipole moments, respectively. Solid lines are obtained from coupled-channel calculations, whereas dashed lines are calculated from the semiclassical model.}%
    \label{fig:NaRb_diff-a}
\end{figure}

NaRb is a particularly interesting case, because the long-range attraction is strong enough to produce a 2-molecule bound state for fields in the center of the shielding region. There are pole-like features in the real part of the scattering length at fields where these bound states enter and leave the well, as shown in Fig.\ \ref{fig:NaRb_diff-a}(a). These poles allow tuning of the scattering length from large negative to large positive values.

\begin{figure}[tbp]
\begin{center}
	\includegraphics[width=0.5\textwidth]{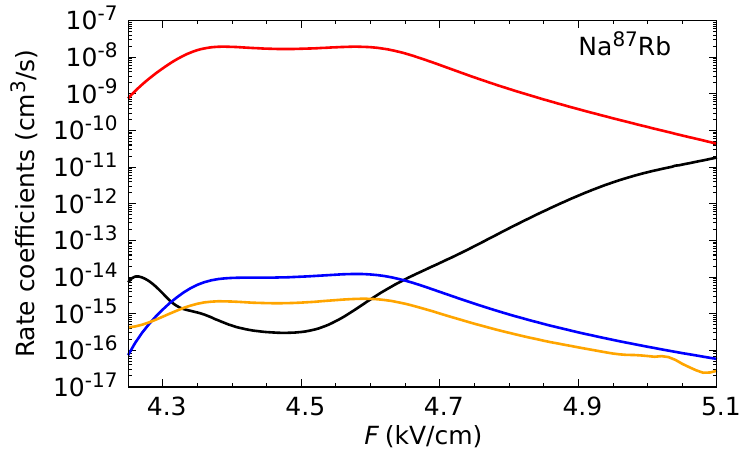}
    \caption{Rate coefficients for elastic scattering (red), and total loss (black) from spin-free coupled-channel calculations on NaRb at collision energy $E_\textrm{coll} = 10$ nK $\times k_\textrm{B}$. The blue and orange curves show state-to-state spin-changing inelastic rate coefficients for 1-molecule inelastic transitions from initial state  $(m_{i,\textrm{Na}}, m_{i,\textrm{Rb}})$ = $(-1/2,1/2)^+$ to $(-1/2,3/2)$ and $(-3/2,3/2)^+$, respectively.}%
    \label{fig:NaRb:rates}
\end{center}
\end{figure}

Figure \ref{fig:NaRb_diff-a}(b) shows $\delta\alpha_{jj}$ for Na$^{87}$Rb (hereafter NaRb) for spin states $j = (-1/2,1/2)^+$ and $(3/2,3/2)$. As for Na$^{39}$K, these states are chosen to give the largest deviations from the spin-free values. For NaRb the fractional deviations in scattering length are somewhat larger than for Na$^{39}$K; they are typically 5\% far from the poles, though necessarily larger close to the poles because the \emph{positions} of the poles are slightly shifted for different $j$. The model of ref.\ \cite{Mukherjee:SU(N):2025} is more accurate for NaRb than for Na$^{39}$K, because there is less cancellation between the positive and negative terms in Eq.\ \ref{eq:dabydD}.

Spin-changing collisions are somewhat faster in NaRb than in Na$^{39}$K, and (for some hyperfine states) provide the dominant source of loss. The resulting rate coefficients are shown in Fig.\ \ref{fig:NaRb:rates} for NaRb in its highest hyperfine state. The inelastic rate nevertheless remains 6 orders of magnitude slower than the elastic rate, so is unlikely to be a problem in experiments to implement SU($N$) symmetry.

\subsection{NaCs}

NaCs is particularly promising for implementing SU($N$) symmetry. It offers very large $N$ (up to 32), and its collisions can be shielded very effectively with either static electric fields \cite{Mukherjee:alkali:2024} or microwave radiation \cite{Bigagli:NaCs:2023}. Both Na and Cs have particularly small nuclear quadrupoles \cite{Aldegunde:singlet:2017}, so that the values of $\delta\alpha_{jj'}$ are expected to be very small \cite{Mukherjee:SU(N):2025}. In static-field shielding, its scattering length can be tuned across a wide range of positive and negative values while maintaining effective shielding \cite{Mukherjee:alkali:2024}.

\begin{figure}[tbp]
\begin{center}
	\includegraphics[width=0.45\textwidth]{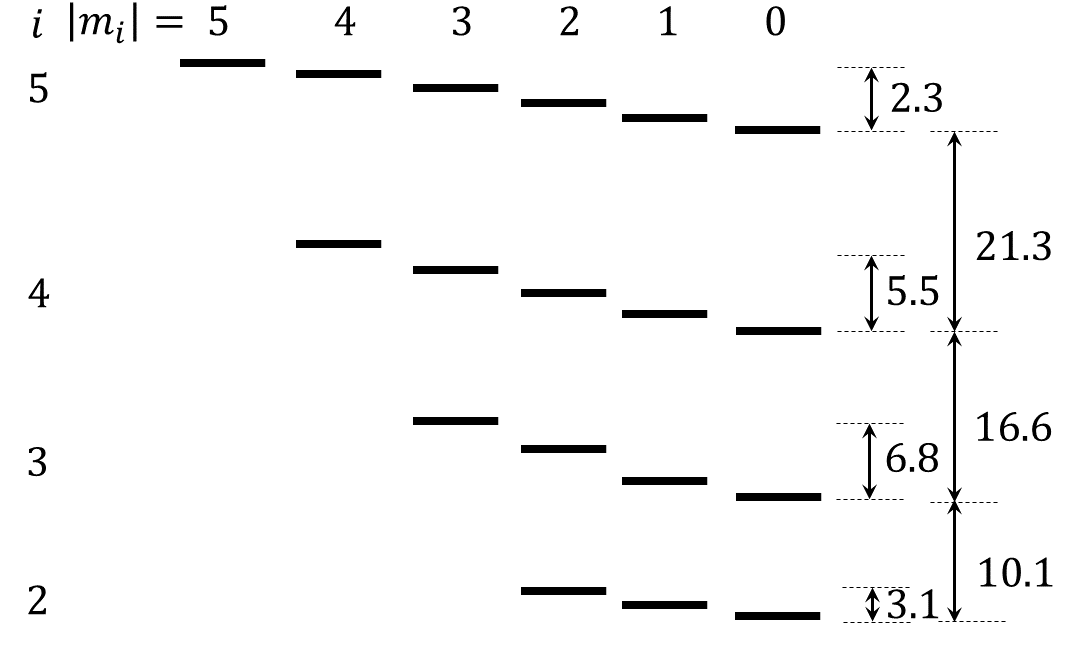}
    \caption{Hyperfine energy levels of NaCs for the rotor level $(\tilde{n},m_n)$ = (1,0) at $F=2.5$ kV/cm. The states are labeled with quantum numbers $i$, the resultant of $i_\textrm{Na}$ and $i_\textrm{Cs}$, and arranged in columns according to $|m_i|$. Spacings (not to scale) are shown in kHz.}%
    \label{fig:NaCs-energy-levels}
\end{center}
\end{figure}

NaCs has a qualitatively different pattern of hyperfine states from NaK and NaRb, shown in Fig.\ \ref{fig:NaCs-energy-levels}. The levels are best described in a coupled representation, with $i$ the resultant of $i_\textrm{Na}$ and $i_\textrm{Cs}$. There are additional splittings because levels with different $m_i$ couple differently to the rotational motion.

\begin{figure}[tbp]
\includegraphics[width=\columnwidth]{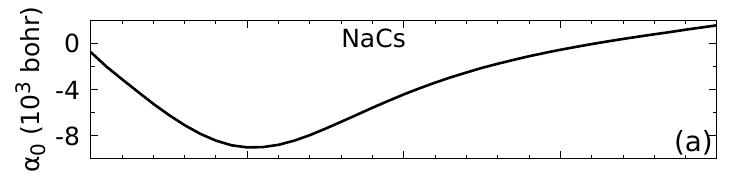}
\includegraphics[width=\columnwidth]{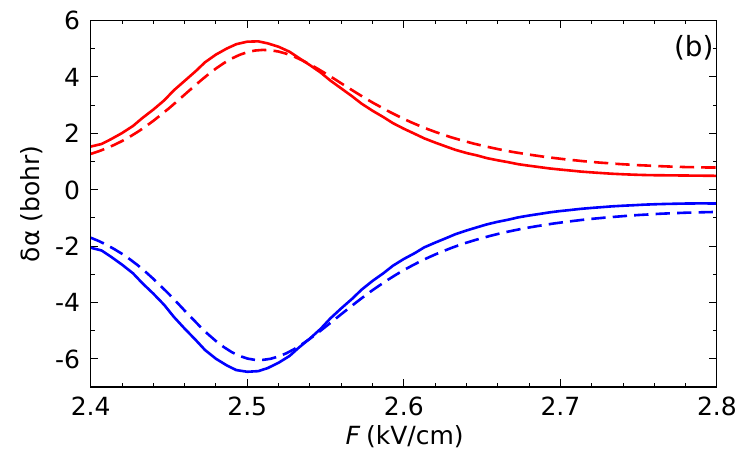}
    \caption{ (a) Real part $\alpha_0(F)$ of the scattering length for NaCs from spin-free calculations. (b) Deviations $\delta\alpha(F)$ from the spin-free value for the combinations of spin states $j+j$ (blue) and $j'+j'$ (red), where $j=(i,m_i)=(3,0)$ and $j'=(3,\pm3)$ . The states $j$ and $j'$ are those with the highest and lowest space-fixed dipole moments, respectively. Solid lines are obtained from coupled-channel calculations, whereas dashed lines are calculated from the semiclassical model.}%
    \label{fig:NaCs_diff-a}
\end{figure}

Because of the large number of spin states, scattering calculations on NaCs with a full spin basis are prohibitively expensive. However, the spin-reduced basis set MFR1 makes the calculations tractable. Figure \ref{fig:NaCs_diff-a} shows values of $\delta\alpha_{jj'}$ for NaCs for collisions between pairs of molecules with the largest and smallest values of $d_j$. The results for other combinations of spin states are expected to lie between these two curves. The range of $\delta\alpha_{jj'}$ is only 12 bohr. For NaCs, $\delta\alpha_{jj'}$ is dominated by the attractive contributions to Eq.\ \ref{eq:Delta-alpha}, and the semiclassical model is accurate to within about 10\% in the region of optimum shielding.

\begin{figure}[tbp]
\begin{center}
	\includegraphics[width=0.5\textwidth]{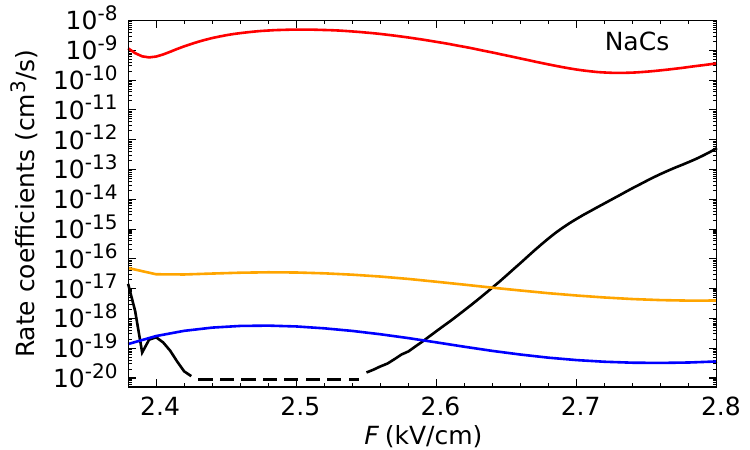}
    \caption{
    Rate coefficients for elastic scattering (red), and total loss (black) from spin-free coupled-channel calculations on NaCs at collision energy $E_\textrm{coll} = 10$ nK $\times k_\textrm{B}$. The blue and orange curves show state-to-state spin-changing inelastic rate coefficients for 1-molecule inelastic transitions from initial state $(i,m_i)$ = (5,0) to (3,1) and (3,0), respectively. The spin-free loss rates at fields between 2.42 and 2.55 kV/cm are limited by numerical precision, and the dashed line shows the upper bound.}%
    \label{fig:NaCs:rates}
\end{center}
\end{figure}

Figure \ref{fig:NaCs:rates} shows the rate coefficients for spin-free elastic scattering and total loss for NaCs, together with the state-to-state rate coefficients for spin-changing collisions from the state  $i=5$, $m_i=0$ to the most important final states. The spin-changing inelastic rates are faster than the spin-free loss rate, but are still low enough to meet the requirements for SU($N$) symmetry. The dominant loss channel involves one molecule relaxing to the state $i=3$, $m_i=0$. This is the lowest state allowed by the selection rule $\Delta i \le 2$ for nuclear quadrupole coupling.

\subsection{Fermionic molecules}

Fermionic molecules behave very similarly to bosonic molecules. For fermions, s-wave collisions are forbidden for pairs of molecules in the same spin state, but are allowed for molecules in different spin states. The s-wave scattering length can be defined only for the latter case, and here the considerations that affect the scattering length are the same as for bosons.

\begin{figure}[tbp]
\includegraphics[width=\columnwidth]{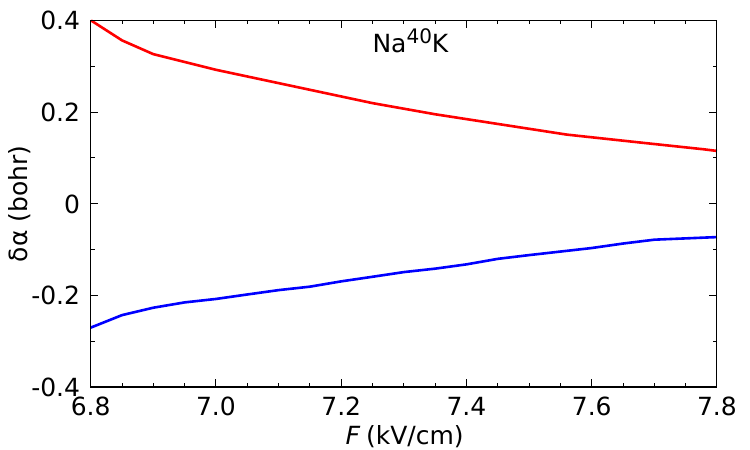}
    \caption{Deviations $\delta\alpha_{jj'}$ in scattering lengths for spin mixtures $j+j^{\prime}$ of fermionic Na$^{40}$K. Here $j+j^{\prime}$ are $(m_{i,\textrm{Na}}, m_{i,^{40}\textrm{K}})=(-1/2,4) + (1/2,4)$ for the blue curve and $(3/2,1)+(3/2,0)$ for the red curve; these are the pairs of states with the highest and lowest space-fixed dipole moments, respectively.}
    \label{fig:Na40K_diff-a}
\end{figure}

Molecules containing $^{40}$K have a very large number of spin states, because of its large nuclear spin, $i_{^{40}\textrm{K}}=4$. Fermionic Na$^{40}$K and $^{40}$K$^{87}$Rb each have 36 hyperfine states, while $^{40}$KCs has 72. Because of this, calculations using a full spin basis are prohibitively expensive. Even calculations with the spin-reduced basis set MFR1 are challenging, but possible. Fig.\ \ref{fig:Na40K_diff-a} shows $\delta\alpha_{jj'}$ for Na$^{40}$K for collisions between pair of states with the largest values of $d_j$, $(m_{i,\textrm{Na}}, m_{i,^{40}\textrm{K}})=(3/2,1)$ and $(3/2,0)$, and states with the smallest values of $d_j$, namely, $(-1/2,4)$ and $(1/2,4)$. For the latter, even the spin-reduced basis set used here includes 8382 functions.  The resulting curves are very similar to those for Na$^{39}$K in Fig.\ \ref{fig:Na39K_diff-a}, except for overall scaling due to the different strength of the quadrupole coupling. The rates of spin-changing collisions are somewhat smaller for Na$^{40}$K than for Na$^{39}$K.
\begin{figure}[tbp]
\begin{center}
	\includegraphics[width=0.5\textwidth]{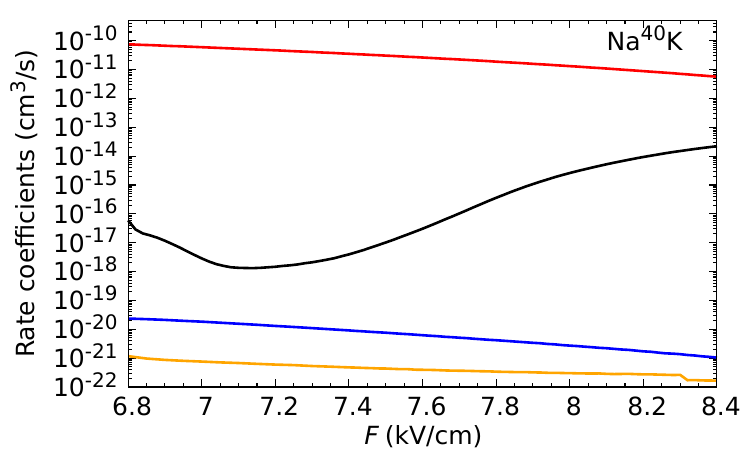}
    \caption{
    Rate coefficients for elastic scattering (red), and total loss (black) from spin-free coupled-channel calculations on Na$^{40}$K at collision energy $E_\textrm{coll} = 10$ nK $\times k_\textrm{B}$. The blue and orange curves show state-to-state spin-changing inelastic rate coefficients for 1-molecule inelastic transitions from initial state $(m_{i,\textrm{Na}}, m_{i,^{40}\textrm{K}}) = (-1/2,4)$ to $(-1/2,3)$ and (1/2,3).}%
    \label{fig:Na40K:rates}
\end{center}
\end{figure}

The properties of collisions of identical molecules are of course quite different for fermions, because they are dominated by p-wave collisions. Figure \ref{fig:Na40K:rates} shows the rate coefficients for spin-free elastic scattering and total loss for Na$^{40}$K, together with the state-to-state rate coefficients for spin-changing collisions from the highest hyperfine spin state $(\tilde{n},m_n,m_{i,\textrm{Na}}, m_{i,^{40}\textrm{K}}) = (1,0,-1/2,4)$. All the inelastic rates are very low. They are potentially important for the creation and lifetime of a single-species Fermi gas, but will make little contribution to losses in mixtures of spin states.

\subsection{Other molecules}

\begin{figure*}[tbp]
	\subfloat[]{
		\includegraphics[width=0.45\textwidth]{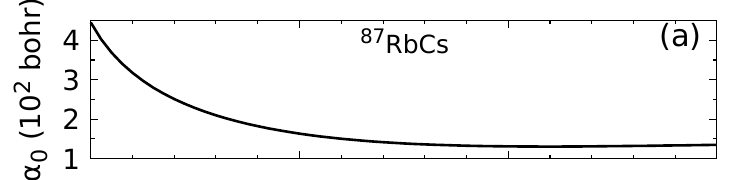}
	}
	\subfloat[]{
		\includegraphics[width=0.45\textwidth]{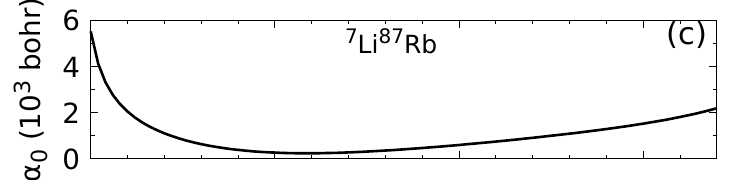}
	}
    \vspace{-1 cm}
	\subfloat[]{
		\includegraphics[width=0.45\textwidth]{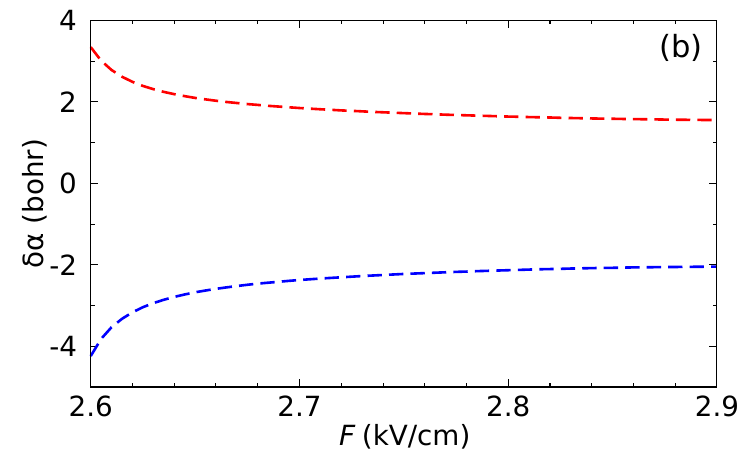}
	}
	\subfloat[]{
		\includegraphics[width=0.45\textwidth]{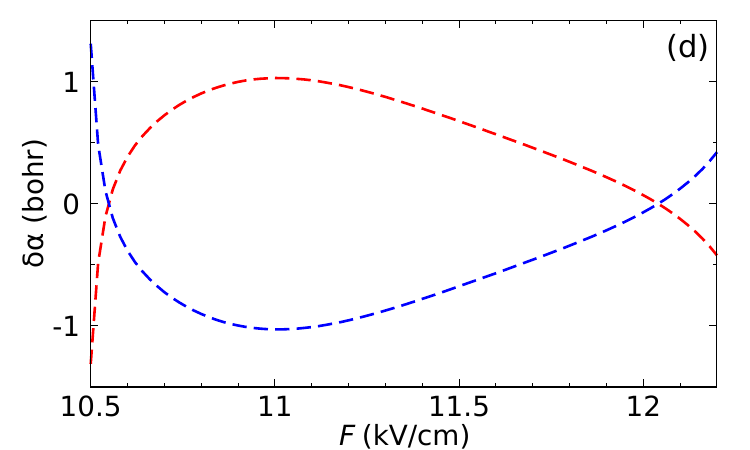}
	}
    \vspace{-0.5 cm}
\caption{Deviations of scattering lengths from their spin-free values for $^{87}$RbCs and $^7$Li$^{87}$Rb from the semiclassical model of Sec.\ \ref{sec:model}. Panels (a) and (c) show the real part $\alpha_0(F)$ of the scattering length obtained from spin-free calculations. Panels (b) and (d) show deviations $\delta \alpha(F)$ from $\alpha_0(F)$ for the combinations of spin states $j+j$ (blue) and $j'+j'$ (red), where $j$ and $j'$ are the states with the highest and lowest space-fixed dipole moments, respectively, at the center of the shielding region. For $^{87}$RbCs, $j=(i,m_i)=(5,\pm5)$ and $j'=(5,0)$; for $^7$Li$^{87}$Rb, $j=(m_{i,^7\textrm{Li}}, m_{i,^{87}\textrm{Rb}})=(\mp1/2, \pm3/2)$ and $j'=(\pm3/2,\pm1/2)$.}
\label{fig:bosonic-mols}
\end{figure*}

The coupled-channel results above validate the semiclassical model for the spin-dependence of scattering lengths, which may thus be used to estimate their magnitude for other similar molecules. The key quantities are the inner turning point of the shielding adiabat and the variation of space-fixed dipole moment between hyperfine states, which were tabulated for alkali dimers of current experimental interest in ref.\ \cite{Mukherjee:SU(N):2025}. The deviations of scattering lengths from their spin-free values are below 1\% for most of the alkali dimers with good shielding, except when the scattering length is very large. Figure \ref{fig:bosonic-mols} shows results from the model for $^{87}$RbCs and $^7$Li$^{87}$Rb. For RbCs, there is again significant cancellation between the positive and negative terms in Eq.\ \ref{eq:dabydD}, and the estimated values of $\Delta\alpha/\alpha_0$ are about $\pm$2\%. For $^7$Li$^{87}$Rb, $\Delta\alpha/\alpha_0$ remains less than 1\%, even though $\alpha_0$ is fairly small ($\sim 270$ bohr at 11 kV/cm).

The spin-changing collisions take place primarily at long-range avoided crossings between an incoming channel with $L=0$ and inelastic channels with $L'=2$ and kinetic-energy release $\Delta E$. These crossings occur outside 1000 bohr for all the systems considered here. They are dominated by couplings due to the larger of the nuclear quadrupole couplings for the two nuclei, $(eQq)_\textrm{max}$. The rates of such collisions are proportional to $m^{5/2} \Delta E^{1/2}$ \cite{Kajita:04a}, where $m$ is the molecular mass and $\Delta E$ scales with $(eQq)_\textrm{max}$. They are also proportional to the square of an effective matrix element $H'$, which itself scales as $\mu^2 (eQq)_\textrm{max}/b$. The overall rates are thus expected to scale between molecules approximately as
\begin{equation}
\frac{m^{5/2} \mu^4 (eQq)_\textrm{max}^{5/2}}{b^2},
\label{eq:ratio}
\end{equation}
where $b$ is the molecular rotational constant. Equation \ref{eq:ratio} predicts overall ratios of rate coefficients for spin-changing collisions that are 1:1000:10 for NaK:NaRb:NaCs. This is in fair agreement with the ratios from coupled-channel calculations, which are approximately 1:5000:20. Further factors can arise from the scattering length, whose effect is neglected in ref.\ \cite{Kajita:04a}. Nevertheless, Eq.\ \ref{eq:ratio} can be used to obtain order-of-magnitude estimates of spin-changing rates for other similar molecules.

\subsection{Effects of magnetic fields}

In the absence of magnetic field, spin states with non-zero $m_f$ exist as degenerate pairs with positive and negative $m_f$. To implement SU$(N$) symmetry, it may be desirable to make these states non-degenerate, which is most simply done with a small magnetic field $B$ parallel to the electric field $F$. Alternatively, it may be convenient to maintain the (typically large) magnetic field used for molecule formation. We have investigated the effect of magnetic fields $B=10$ and 500~G for Na$^{39}$K, Na$^{87}$Rb and NaCs at the center of the shielding region, using basis set MFR1. We find that such magnetic fields have little effect on the ranges of scattering lengths for NaK and NaRb, but the larger field increases the range by about a factor of 2 for NaCs. This arises because such a field substantially changes the nature of the monomer states for NaCs.

\section{Conclusions}
\label{sec:conclusions}

We have shown that ultracold alkali dimers that are shielded from destructive collisions with static electric fields are very promising systems for implementing SU($N$) symmetry. The calculated scattering lengths are very similar for all pairs of spin states. Spin-changing collisions are very slow, as are total loss rates. The alkali dimers offer several advantages over existing implementations of SU($N$) symmetry using alkaline-earth-like atoms. In particular, alkali dimers offer larger $N$, up to 36 for the ultracold molecules studied in current experiments. There are both bosonic and fermionic molecules available, and the interactions can be tuned to be either attractive or repulsive.

We have carried out coupled-channel calculations on shielded collisions, including the full hyperfine structure of the colliding molecules. To achieve this, we have developed special methods to accommodate the enormous basis sets involved. In particular, we use Van Vleck transformations to reduce the number of basis functions included explicitly in the coupled-channel basis sets, while taking account of functions outside this space perturbatively. We have also developed spin-restricted basis sets and shown that they accurately reproduce calculations with full spin basis sets.

We have carried out calculations on the representative molecules Na$^{39}$K, Na$^{40}$K, Na$^{87}$Rb and NaCs. Na$^{40}$K and NaCs are particularly challenging because of the large spins of the $^{40}$K and Cs nuclei. For all systems, the scattering lengths for different spin states are within 1\% of the spin-free values, except for NaRb, where the scattering length is very large. We expect the same to be true for other alkali dimers with effective shielding. We have developed and tested a semiclassical model of the spin-dependence, based on space-fixed dipole moments of individual molecules and spin-free scattering calculations, and shown that it gives good estimates of the variation in scattering lengths. We have used the model to give estimates of the spin-dependence for $^{87}$RbCs and $^7$Li$^{87}$Rb.

The different molecules studied have significantly different patterns of hyperfine states. This is manifested in different selection rules for spin-changing collisions, and rates of such collisions that differ by a factor of 5000 between Na$^{39}$K and Na$^{87}$Rb. Nevertheless, the rates of spin-changing collisions are at least 6 orders of magnitude lower than the rates of elastic collisions for all the systems studied. We have presented a scaling law that qualitatively explains the variation in spin-changing rates between systems.

We have investigated the effects of magnetic fields parallel to the electric field, and found that the deviations in scattering lengths depend only weakly on them.

Current experiments on ultracold alkali dimers often use shielding with microwave fields instead of static electric fields. The physics of microwave shielding has strong similarities to that of static-field shielding, so we anticipate that SU($N$) symmetry can be achieved in this case too. However, further work is needed to quantify this.

Realization of SU($N$) symmetry with alkali dimers will open the door to study much new physics. Systems with attractive interactions will allow experiments to explore the formation and ordering of energetically favorable clusters~\cite{Rapp:2007, Inaba:2009, Titvinidze:2011, Pohlmann:2013, Xu:SU3:2023}. For fermionic systems, SU($N$) symmetry will enhance quantum fluctuations and topological order \cite{Hermele:2009, Chen:multiflavor:2024}. It may allow even lower temperatures than so far achieved with alkaline-earth atoms \cite{Taie:2022}. For bosonic systems, SU($N$) symmetry will provide opportunities to study the phases of exotic ferromagnets~\cite{Polychronakos:2023}. There are ultracold molecules available in current experiments with properties that offer access to all these domains.

\section*{Rights retention statement}

For the purpose of open access, the authors have applied a Creative Commons Attribution (CC BY) licence to any Author Accepted Manuscript version arising from this submission.

\section*{Data availability statement}

Data supporting this study are openly available from Durham University \cite{DOI_data-AlkSU_N}.

\begin{acknowledgements}
We are grateful to Kaden Hazzard for interesting us in this topic, and to James Croft, Matthew Frye and Ruth Le Sueur for valuable discussions. We acknowledge support from the U.K. Engineering and Physical Sciences Research Council (EPSRC) Grant Nos.\ EP/W00299X/1, and EP/V011677/1.
\end{acknowledgements}

\newpage
\bibliographystyle{../long_bib}
\bibliography{../all,AlkSU_N_data}

\end{document}